\documentclass[prl,twocolumn,superscriptaddress,showpacs]{revtex4}
\usepackage{verbatim}
\usepackage{graphicx}
\graphicspath{{../}{./figures/}{../figures/}}
\usepackage{amssymb,amsmath}
\usepackage{color}
\definecolor{refblue}{rgb}{0.,.17,.45}
\usepackage[
            pdfstartview=FitH,
            bookmarks=false,
            colorlinks=true,
            citecolor=blue,
            linkcolor=blue]
            {hyperref}

\def\nuc#1#2{\relax\ifmmode{}^{#1}{\protect\text{#2}}\else${}^{#1}$#2\fi}
\newcommand{\betadecay}{
  \mbox{$\longrightarrow$
        \mbox{\hspace{-1.3em}}\raisebox{+0.9ex}{\scriptsize $\beta$}
        \makebox[0.2em]{}
}}
\DeclareSymbolFont{symbols}{OMS}{cmsy}{m}{n}

\begin{document}

\bibliographystyle{prsty}

\title{Beta-delayed deuteron emission from $^\mathbf{11}$Li:
decay of the halo}

\author{R.~Raabe}
\affiliation{Instituut voor Kern- en Stralingsfysica, K.U.Leuven,
B-3001 Leuven, Belgium}
\author{A.~Andreyev}
\affiliation{TRIUMF, Vancouver, British Columbia, Canada V6T 2A3}
\author{M.J.G.~Borge}
\affiliation{Instituto de Estructura de la Materia, CSIC, Madrid,
Spain}
\author{L.~Buchmann}
\affiliation{TRIUMF, Vancouver, British Columbia, Canada V6T 2A3}
\author{P.~Capel}
\affiliation{TRIUMF, Vancouver, British Columbia, Canada V6T 2A3}
\author{H.O.U.~Fynbo}
\affiliation{Department of Physics and Astronomy, University of
Aarhus, DK-8000 Aarhus C, Denmark}
\author{M.~Huyse}
\affiliation{Instituut voor Kern- en Stralingsfysica, K.U.Leuven,
B-3001 Leuven, Belgium} \affiliation{TRIUMF, Vancouver, British
Columbia, Canada V6T 2A3}
\author{R.~Kanungo}
\affiliation{TRIUMF, Vancouver, British Columbia, Canada V6T 2A3}
\author{T.~Kirchner}
\affiliation{TRIUMF, Vancouver, British Columbia, Canada V6T 2A3}
\author{C.~Mattoon}
\affiliation{Department of Physics, Colorado School of Mines,
Golden, Colorado 80401, USA}
\author{A.C.~Morton}
\affiliation{TRIUMF, Vancouver, British Columbia, Canada V6T 2A3}
\author{I.~Mukha}
\affiliation{Instituut voor Kern- en Stralingsfysica, K.U.Leuven,
B-3001 Leuven, Belgium} \affiliation{Universidad de Sevilla,
ES-41080 Seville, Spain}
\author{J.~Pearson}
\affiliation{TRIUMF, Vancouver, British Columbia, Canada V6T 2A3}
\author{J.~Ponsaers}
\affiliation{Instituut voor Kern- en Stralingsfysica, K.U.Leuven,
B-3001 Leuven, Belgium}
\author{J.J.~Ressler}
\affiliation{Department of Chemistry, Simon Fraser University,
Burnaby, B.C. Canada V5A-1S6} \affiliation{TRIUMF, Vancouver,
British Columbia, Canada V6T 2A3}
\author{K.~Riisager}
\affiliation{Department of Physics and Astronomy, University of
Aarhus, DK-8000 Aarhus C, Denmark}
\author{C.~Ruiz}
\affiliation{Department of Physics, Simon Fraser University,
Burnaby, B.C. Canada V5A-1S6} \affiliation{TRIUMF, Vancouver,
British Columbia, Canada V6T 2A3}
\author{G.~Ruprecht}
\affiliation{TRIUMF, Vancouver, British Columbia, Canada V6T 2A3}
\author{F.~Sarazin}
\affiliation{Department of Physics, Colorado School of Mines,
Golden, Colorado 80401, USA}
\author{O.~Tengblad}
\affiliation{Instituto de Estructura de la Materia, CSIC, Madrid,
Spain}
\author{P.~Van~Duppen}
\affiliation{Instituut voor Kern- en Stralingsfysica, K.U.Leuven,
B-3001 Leuven, Belgium}
\author{P.~Walden}
\affiliation{TRIUMF, Vancouver, British Columbia, Canada V6T 2A3}

\date{\today}

\begin{abstract}

The deuteron-emission channel in the $\beta$-decay of the
halo-nucleus \nuc{11}{Li} was measured at the ISAC facility at
TRIUMF by implanting post-accelerated \nuc{11}{Li} ions into a
segmented silicon detector. The events of interest were identified
by correlating the decays of \nuc{11}{Li} with those of the
daughter nuclei. This method allowed the energy spectrum of the
emitted deuterons to be extracted, free from contributions from
other channels, and
a precise value for the branching ratio $B_d = 1.30(13) \times
10^{-4}$ to be deduced for $E_\mathrm{c.m.} >
200$ keV.
The results provide the first unambiguous experimental evidence that the
decay takes place essentially in the halo of
\nuc{11}{Li}, and that it proceeds mainly to the
\nuc{9}{Li} + $d$ continuum, opening up a new means to 	
study of the halo 
wave function of \nuc{11}{Li}.
\end{abstract}

\pacs{23.40.-s; 23.40.Hc; 27.20.+n}

\maketitle

The nuclear halo \cite{Han95} is among the most peculiar features
discovered in unstable nuclei. 
The \nuc{11}{Li} nucleus is the showcase of a two-neutron halo
system, with its very extended matter distribution 
related to the small energy necessary to remove the neutrons,
$S_\mathrm{2n} = 378(5)$ keV \cite{Bac08}. 
Considerable effort has been expended to determine the characteristics of
this unstable, short-lived nucleus ($T_{1/2} = 8.5(2)$ ms
\cite{Ajz90}).
Among the available probes, the $\beta$-decay has the advantage of being
described by a well-established theory, and thus provides a valuable
tool for the investigation of the properties of the ground state of
the parent nucleus.

In the case of \nuc{11}{Li}, one decay channel is of special
interest: the $\beta$-delayed deuteron emission
\nuc{11}{Li}\betadecay\nuc{9}{Li} + $d$.
This mode is related to the possibility that 
in halo nuclei
the core and the halo particles could decay,
more or less independently, into different channels
\cite{Nil00}.
Evidence of the $\beta$-decay of the \nuc{9}{Li} core in
\nuc{11}{Li} with survival of the two-neutron halo was reported
%
in Ref. \cite{Sar04}. On the other hand, according to
calculations \cite{Ohb95,Zhu95,Bay06} the deuteron-emission channel
should be dominated by the decay of a neutron in the halo: a
``halo decay''. 
This 
may be measured via the \nuc{11}{Li}\betadecay\nuc{9}{Li} + $d$ decay
probability, or the branching ratio $B_d$.
In addition, the $\beta$-delayed deuteron decay of \nuc{11}{Li}
could proceed directly to the continuum, without forming an
intermediate state (resonance) in the daughter nucleus
\nuc{11}{Be}. The information on the initial \nuc{11}{Li} wave
function would then be more easily accessible. Both $B_d$
and the energy distribution of the emitted deuterons are
related to this. 
%


We 
report here the first measurement of the relevant quantities, $B_d$ and
the energy of the emitted \nuc{9}{Li} and $d$ ions,
without contributions from other channels,
by studying the decay correlations of a post-accelerated \nuc{11}{Li}
beam implanted in a segmented silicon 
detector \cite{Pau95,Smi05}.


The $\beta$-decay of \nuc{11}{Li} is complex. The large mass
difference between \nuc{11}{Li} and its daughter \nuc{11}{Be}
($Q = 20.6$ MeV) implies that many decay channels to bound and
unbound states in \nuc{11}{Be} are open. In the latter cases, the
daughter breaks into fragments, and emission of one \cite{Roe74}, two
\cite{Azu79}, and three neutrons \cite{Azu80}, $\alpha$ particles
and \nuc{6}{He} \cite{Lan81}, tritons \cite{Lan84a} and deuterons
\cite{Muk96} has been observed.
A summary 
is given in Table \ref{tab:11li_decay}.
\begin{table}
 \renewcommand{\arraystretch}{0.7}
 \caption{$\beta$-decay channels of \nuc{11}{Li} and decay
 characteristics of its unstable daughters.
 Branching ratios for the
 \nuc{11}{Li} channels are estimates from previous works.}
 \label{tab:11li_decay}
 \begin{footnotesize}
 \begin{ruledtabular}
  \begin{tabular*}{\hsize}{lccllll}
   \multicolumn{5}{l}{Decay channel} &
   \multicolumn{2}{l}{Daughter half life and decay}\\
   \hline \\ [-2.5ex]
   (\emph{a}) & 6.3\% & \cite{Bor97} & \nuc{11}{Be} + $\gamma$ & &
   13.81 s, & $\left\{\begin{array}{l}
              \hspace{-.5em}(1)\ \nuc{11}{B} + \gamma,\ 93(1)\%\footnotemark[1]\\
     [-0.2ex] \hspace{-.5em}(2)\ \nuc{7}{Li} + \alpha,\ 7(1)\%\footnotemark[1]\\
                      \end{array}
               \right.$ \\
   (\emph{b}) & 87.6\% & \cite{Bor97} & \nuc{10}{Be} + $n$ & &
   $10^6$ y & \\
   (\emph{c}) & 4.2\% & \cite{Bor97} & \nuc{9}{Be} + 2$n$ & &
   stable & \\
   (\emph{d}) & 1.0\% & \cite{Lan84} & \nuc{6}{He} + $\alpha$ + $n$ & &
   807 ms, & \nuc{6}{Li} g.s., $\approx$100\%\\
   (\emph{e}) & 1.9\% & \cite{Bor97} & $2\alpha$ + 3$n$ & &
   stable & \\
   (\emph{f}) & 0.01\% & \cite{Muk96} & \nuc{8}{Li} + $t$ & &
   838 ms, & $2\alpha$, 100\%\\
   (\emph{g}) & 0.01\% & \cite{Muk96} & \nuc{9}{Li} + $d$ & &
   178 ms, & $\left\{\begin{array}{l}
              \hspace{-.5em}(1)\ \nuc{9}{Be}\ \mathrm{g.s.},\ 49.2(9)\%\footnotemark[2]\\ 
     [-0.2ex] \hspace{-.5em}(2)\ 2\alpha + n, 50.8(9)\%\footnotemark[2]\\ 
                    \end{array}
              \right.$ \\
  \end{tabular*}
 \end{ruledtabular}
 \end{footnotesize}
 \footnotetext[1]{From this work, assuming 6.3(6)\% \cite{Bor97}
 as branching ratio for 
 (\emph{a}).}
 \footnotetext[2]{From \cite{Til04}.}
\end{table}
The $\beta$-delayed emission of deuterons has also been observed in
the decay of the two-neutron halo nucleus \nuc{6}{He}
\cite{Rii90,Bor93,Ant02,Smi05}.
However, the \nuc{6}{He}\betadecay$\alpha$ + $d$ channel has
contributions from both the halo and core parts of the \nuc{6}{He}
ground-state wave function \cite{Bay94,Cso94,Tur06}; the two
contributions interfere destructively leading to a very small value
of the corresponding branching ratio of about $10^{-6}$. In
\nuc{11}{Li}, on the other hand,
the long-range behaviour of the wave function, which is more extended
than in \nuc{6}{He}, could favour the unperturbed decay of the
halo particles and the suppression of the core contribution,
producing a 
%
%
%
comparatively larger branching ratio
of the order of $10^{-4}$ \cite{Ohb95,Zhu95,Bay06}.

The experimental detection of the deuterons from the
\nuc{11}{Li}\betadecay\nuc{9}{Li} + $d$ decay is complicated by
their low energy -- the \nuc{9}{Li} + $d$ threshold in
\nuc{11}{Be} is at $E^* = 17.9$ MeV, thus the energy available in
this channel is $Q_d = 2.7$ MeV.
In addition, deuterons are difficult to separate from the tritons
emitted in the \nuc{8}{Li} + $t$ channel ($Q_t = 4.9$ MeV).
%
%
The only previous work reporting their detection 
is that by Mukha \emph{et al.} \cite{Muk96}.
The authors measured a cumulative spectrum for $Z = 1$ particles
from the activity of \nuc{11}{Li} nuclei deposited on a thin foil,
%
%
%
and estimated a branching ratio $B \gtrsim 10^{-4}$.
Evidence of the actual presence of deuterons was provided by the
correlated detection of a few decay events of the associated
nucleus \nuc{9}{Li}.

In the present work
we used a different experimental technique and directly implanted
the \nuc{11}{Li} nuclei
in a thin, highly segmented silicon detector. 
The decay channels were identified through the time and position
correlation between the implanted nuclei and subsequent parent and
daughter decays.
%


The \nuc{11}{Li} nuclei were produced
by bombarding a thick Ta target with the 500 MeV, 35 $\mu$A proton
beam from the TRIUMF Cyclotron. After extraction and selection by a
magnetic analyzer, the \nuc{11}{Li} nuclei were post-accelerated to
an energy of 1.5 MeV/nucleon in the ISAC-I accelerator \cite{Dau03}.
%
%
The \nuc{11}{Li} implantation into
the detector was continuous;
to preserve the correlation between an implantation event and the
subsequent decays, we used only a fraction of the available
intensity (about 200 particles per second of the total few
thousands).
This also helped reduce the dead time of the acquisition
system. 
%
The beam was de-focused to spread the implantations on the whole
detector surface.
About $85 \times 10^6$ \nuc{11}{Li} ions were implanted in 130
hours of beam time.

The double-sided silicon strip detector was 78 $\mu$m thick, with
an active area of 16$\times$16 mm$^2$. The 48 strips on each side
(300 $\mu$m wide, with those on the back face perpendicular to
those on the front) defined a total of 2304 pixels. The efficiency
for the detection of low-energy events,
measured using pulsed signals of calibrated amplitude, was about
75\% at 200 keV, rising to 100\% at 300 keV.
For all events in each pixel, the recorded information
included the deposited energy and a timestamp.
%
%
%
%
The main characteristics of the method have been described in Ref.
\cite{Smi05}.
%
A precise normalization is ensured by the direct correlation
between implanted ions and detected decays. By selecting
single-pixel events (only one signal in each set of strips),
$\beta$ particles are strongly suppressed since their specific
energy deposition is low -- less than 1\% of all $\beta$'s are
detected above the threshold, and less than one in 10$^6$ deposit
more than 600 keV \cite{Smi05}.
Conversely, the energy deposited by ions in virtually all
\nuc{9}{Li} + $d$ decay events is completely collected in a single
pixel.
%
%
%
The range of 1 MeV deuterons in silicon is 12.2 $\mu$m and the
interstrip distance was 35 $\mu$m. The implantation depth of the
\nuc{11}{Li} was 43(2) $\mu$m according to SRIM \cite{Zie85}
calculations; this results in 99.9\% of the \nuc{9}{Li} + $d$
events being contained in the implantation pixel \cite{Smi05}.
%
%
%

\begin{figure}
 \includegraphics[width=.48\textwidth]{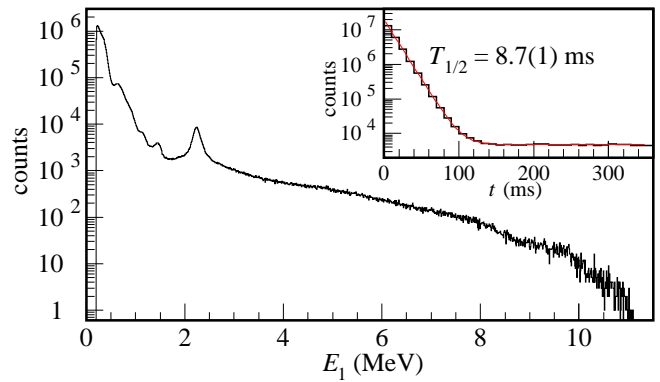}
\caption{(Color online) Energy spectrum of ($E_1,t_1$) 
events, defined
as those following a \nuc{11}{Li} implantation in the same pixel
within 40 ms: these are \nuc{11}{Li} decays. The insert shows the
time behaviour of all events following an implantation:
the short-time component has the expected \nuc{11}{Li} half
life.}
 \label{fig:11li_decay}
\end{figure}

Fig. \ref{fig:11li_decay} shows the events immediately following a
\nuc{11}{Li} implantation in the same pixel. The time behavior has
a short component with a half life $T_{1/2} = 8.7(1)$ ms as
expected for \nuc{11}{Li} decays. The energy spectrum of events
within the first 40 ms -- ($E_1,t_1$) 
events -- contains less than 0.1\% background from uncorrelated
activity.
%
%
The features of this \nuc{11}{Li} $\beta$-decay spectrum will be
discussed in a separate publication.
%
%

\begin{figure}[t]
 \includegraphics[width=.48\textwidth]{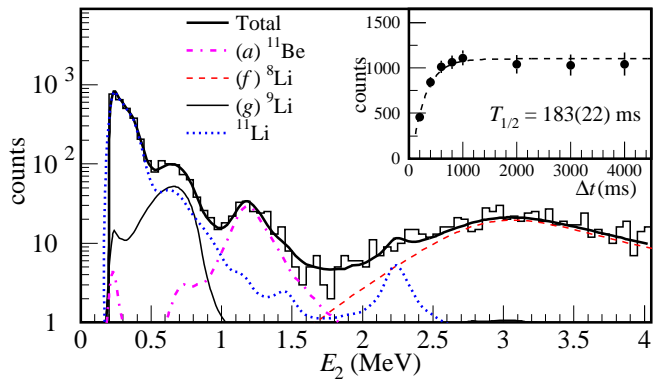}
\caption{(Color online) Example of a fit of ($E_2,t_2$) 
events, with contributions from daughter decays in channels (\emph{a}),
(\emph{f}) and (\emph{g}), and uncorrelated \nuc{11}{Li} decays.
This spectrum was obtained for 
$t_2 - t_1 < 200$ ms and for
$E_1$ 
between 0.5 MeV and 2.0 MeV.
The insert shows the number of \nuc{9}{Li} decay events in the
($E_2,t_2$)
spectra for $t_2 - t_1 < \Delta t$ as function of $\Delta t$. The
fitted time behaviour is in agreement with the expected \nuc{9}{Li}
half life, $T_{1/2} = 178.3(4)$ ms \cite{Til04}.}
 \label{fig:e2_fit}
\end{figure}

($E_2,t_2$) events were defined as the first decay events following
the \nuc{11}{Li} decays ($E_1,t_1$) 
in the same pixel, before a subsequent implantation 
took place. Such daughter-decay events were used to identify the
decay channels of interest (Fig. \ref{fig:e2_fit}).
%
%
%
%
%
Referring to Table
\ref{tab:11li_decay}, and taking into account that pure
$\beta$-emission (or $\beta$-$\gamma$) decays are strongly
suppressed in our setup, we expect contributions
mainly from the daughters in channels (\emph{f}) and (\emph{g}).
The considerably shorter halflife of the \nuc{8,9}{Li} daughters
compared to the average time between two implantations in the same
pixel (about 8.5 s) shows that these daughter decays are truly correlated
with the previous \nuc{11}{Li} decay. Daughter decays from
channel (\emph{a}) also appear, but they are uncorrelated because their 
\nuc{11}{Li} mother decay branch is mostly not
detected (pure $\beta$-$\gamma$), and due to the long half life
of \nuc{11}{Be}.
%
Finally, the spectrum of ($E_2,t_2$) contains 
uncorrelated decays of \nuc{11}{Li}, due to undetected
implantations taking place between the decays at $t_1$ and $t_2$.
The undetected implantations (mainly caused by the dead time of the
acquisition system) are only a few percent of the total; still,
because of the small branching ratios of the other channels, at low
energy the uncorrelated \nuc{11}{Li} decays 
dominate the spectrum of ($E_2,t_2$) 
events.

To correctly evaluate each contribution, the decays of the
daughter channels of interest were separately studied in dedicated
measurements with our
setup. Beams of \nuc{9}{Li} and \nuc{8}{Li} ions were produced at
ISAC and implanted in the segmented detector
at depths (close to the detector middle plane)
to ensure that the
ions emitted in the decay would not escape.
%
Beam modulations of 0.5 s ``beam on'' - 0.5 s ``beam off''
and 2.5 s on - 2.5 s off were used for \nuc{9}{Li} and \nuc{8}{Li}
respectively.
%
%
For the decay
of \nuc{11}{Be} we used again a \nuc{11}{Li} beam, with a
modulation 20 s on - 20 s off: a short time after stopping the
beam, only the long-living \nuc{11}{Be} activity remained.
%
The shapes of the collected spectra differed from the
ones from literature, 
reflecting the fact that the sum energy of all ions emitted in
each decay was measured in our setup.
These measurements also allowed the efficiency for the
detection of each decay in our setup to be determined.
In particular, the efficiency
for the decay of \nuc{8}{Li} was 100(1)\% (equal to the branching
ratio reported in Table \ref{tab:11li_decay}) because of the high
energy of the emitted $\alpha$ particles; for \nuc{9}{Li}, it was
only 36(1)\%,
resulting from the convolution of the spectrum of the low-energy
$\alpha$ particles from the decay with our detection efficiency at
energies below 300 keV.

\begin{figure}
 \includegraphics[width=.48\textwidth]{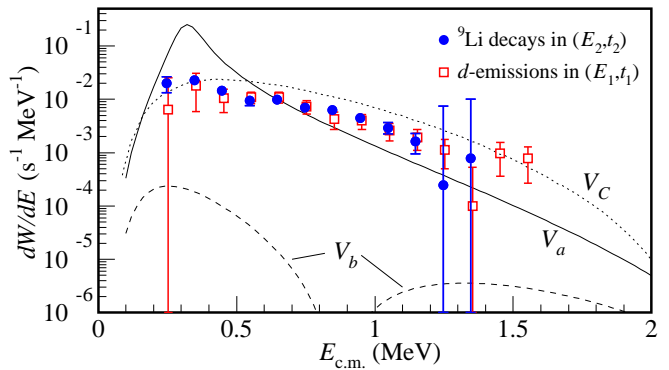}
\caption{(Color online) Transition probability
for the \nuc{11}{Li}{\protect \betadecay} \nuc{9}{Li} + $d$
decay channel.
The procedures to obtain the experimental data are
explained in the text.
The curves are predictions from Refs. \cite{Zhu95} ($V_C$) and
\cite{Bay06} ($V_a$, $V_b$).}
 \label{fig:dW/dE}
\end{figure}

We adopted two procedures to extract the branching ratio and the
energy spectrum of the \nuc{11}{Li}\betadecay\nuc{9}{Li} + $d$
channel. The first consisted of counting the \nuc{9}{Li} decays
among the ($E_2,t_2$) 
events. For this purpose, we fitted the ($E_2,t_2$) 
spectrum with the measured spectra of the daughter decays in the
channels (\emph{a}), (\emph{f}) and (\emph{g}) (the one of
interest), and the spectrum of \nuc{11}{Li} decays (from the
undetected implantations).
The fit was repeated for different time windows
$\Delta t$ (with $t_2 - t_1 < \Delta t$)
and various energies $E_1$ of the first decay; 
an example is shown in Fig. \ref{fig:e2_fit}.
The asymptotic value of the number of \nuc{9}{Li} decays for
$\Delta t \rightarrow +\infty$ (about 3200 for the whole energy
range in $E_1$), 
corrected for the detection efficiency, yields the number of
deuteron-emission events.
An additional small correction was applied to
account for the possibility that a
subsequent implantation event took place in the same pixel before
the \nuc{9}{Li} decay.
By repeating the procedure for different energy intervals in $E_1$,
a differential branching ratio was obtained or, in other words, the
decay probability $dW/dE$ as function of the total \nuc{9}{Li}-$d$
energy $E_\mathrm{c.m.}$. The result is shown in Fig.
\ref{fig:dW/dE} (filled circles); the uncertainties (statistical
only are plotted) propagate from the errors in the fit
of the ($E_2,t_2$) 
spectra.

The second approach consisted of directly selecting
deuteron-emission events among the ($E_1,t_1$)
ones by applying conditions on the subsequent ($E_2,t_2$) 
events in order to maximize the number of \nuc{9}{Li} decays.
This was 
achieved by requiring $t_2 - t_1 < 200$ ms and
$E_2$ 
between 0.55 MeV and 0.8 MeV (see also Fig. \ref{fig:e2_fit}).
The amount of ``background'' 
present in the selection was determined from the fit of the
($E_2,t_2$) 
spectrum; after subtraction, about 700 \nuc{9}{Li} + $d$ events
remained in the ($E_1,t_1$) 
spectrum.
The normalization took into account the detection efficiency for
the \nuc{9}{Li} decay events, plus the factors introduced by the
narrow selection of the ($E_2,t_2$) 
events. The resulting spectrum is shown in Fig. \ref{fig:dW/dE}
(hollow squares); because of the smaller statistics, the
uncertainties are larger than for the first method.

The two methods produced consistent results within the statistical
uncertainties. An additional 8\% systematic error is present in
both cases, related to the overall normalization.
For the total
branching ratio of the deuteron-emission channel, the first method
(with the better statistics) yields $B_d = 1.30(13) \times 10^{-4}$
for a \nuc{9}{Li}-$d$ total energy $E_\mathrm{c.m.} > 200$ keV. The
value obtained with the second method is $B_d = 1.08(23) \times
10^{-4}$.

$B_d$ is sensitive to two aspects:
the decay of
the halo part of the \nuc{11}{Li} wavefunction,
and
the decay
directly to the continuum rather than to a
resonance in the daughter nucleus \nuc{11}{Be}. Concerning the
first aspect, we have already pointed out how contributions from the
core, in the case of \nuc{6}{He}, interfere destructively with
those from the halo, reducing the branching ratio to about
$10^{-6}$. As for the second aspect, in the calculations published
so far \cite{Ohb95,Zhu95,Bay06}, it has been shown that a maximum
value of $B_d \approx 10^{-4}$ could only be reduced (by as much as
two orders of magnitude) if a resonance existed below the
\nuc{9}{Li} + $d$ threshold in \nuc{11}{Be}. We conclude that the
value which we measured confirms that the deuteron-emission decay
takes place essentially in the halo of \nuc{11}{Li}, without significant
interference of contributions from the decay of the \nuc{9}{Li}
core
(which was in turn
observed in other channels of the \nuc{11}{Li} decay
\cite{Sar04}).

The question of whether the decay proceeds directly to the
continuum or through a resonance is also related to the
energy distribution
of the emitted ions, Fig. \ref{fig:dW/dE}. In Refs.
\cite{Ohb95,Zhu95,Bay06}, different potentials for the
\nuc{9}{Li}-$d$ interaction were used to reproduce a resonance
around the \nuc{9}{Li} + $d$ threshold in \nuc{11}{Be}. The curves
in Fig. \ref{fig:dW/dE} are taken from those studies and are
representative of the results. The curves labeled $V_a$ and $V_b$
(from Ref. \cite{Bay06}) correspond to potentials producing a
resonance, respectively 0.33 MeV above and 0.18 MeV below the
\nuc{9}{Li} + $d$ threshold; the one labeled $V_C$ (from Ref.
\cite{Zhu95}) corresponds to the Coulomb potential only, without a
resonance. The integral value of the transition probability rules
out the $V_b$ case. 
A resonance above the threshold, curve $V_a$,
generates a pronounced maximum, shifting the spectrum towards lower
energies and reducing it at energies above the resonance; absorption 
into other decay channels \cite{Bay06} may reduce the height of the
maximum but would not affect the spectrum above 1 MeV.
%
This, and the fact that our data are better reproduced by the
curve $V_C$, indicate that we do not see a resonance in our observation
window or just below it.
We recall here that
a resonance in \nuc{11}{Be} has been observed at an excitation
energy $E^* \approx 18.1$ MeV by Borge \emph{et al.} \cite{Bor97a},
about 200 keV above the \nuc{9}{Li} + $d$ threshold. Their
analysis, however, based on a branching ratio $B_d \approx 10^{-4}$,
showed that it is unlikely that the deuteron-emission decay takes place
through this level. Our results thus support the picture of a decay
proceeding mainly to the continuum. This implies that the decay
matrix elements represent a Fourier transform of the \nuc{11}{Li}
ground-state wave function 
to which our data should then be very sensitive.
These aspects should be addressed in future theoretical studies.


In conclusion, we have measured the deuteron-emission channel in the
$\beta$-decay of \nuc{11}{Li} by implanting a post-accelerated beam
of \nuc{11}{Li} ions into a highly-segmented silicon detector and
identifying the channel of interest by time and position
correlation of implantation events and subsequent parent-daughter
decays. Precision data were obtained for the branching ratio and
the \nuc{9}{Li}-$d$ total energy spectrum down to an energy
threshold of 200 keV. The results provide the first clear experimental
support for the deuteron-emission decay taking place essentially in
the halo of \nuc{11}{Li} and indicate that the decay proceeds
mainly to the \nuc{9}{Li} + $d$ continuum, opening up the possibility
for a detailed study of the halo 
wave function
of \nuc{11}{Li} complementary to those based on nuclear reactions.





\begin{acknowledgments}

We thank R.E. Laxdal, P. Bricault and the operators of the ISAC
facility at TRIUMF for their efforts to produce the Li
beams. We also thank D. Baye, P. Descouvemont and
M. Zhukov for instructive discussions. 
R.R. is a
Postdoctoral Fellow of the Research Foundation - Flanders (FWO).
This work was supported by the BriX Interuniversity Attraction Poles
Programme - Belgian Science Policy (IUAP) under project P6/23 and
the Research Foundation - Flanders (FWO), Belgium.

\end{acknowledgments}
\vspace{-.3cm}

\end{document}